\documentclass[magnetism,article,accept,pdftex,moreauthors]{Definitions/mdpi} 
\firstpage{1} 
\pubvolume{1}
\issuenum{1}
\articlenumber{0}
\pubyear{2025}
\copyrightyear{2025}
\externaleditor{Firstname Lastname} 
\datereceived{26 January 2025} 
\daterevised{2 April 2025} 
\dateaccepted{16 April 2025} 
\datepublished{ } 
\hreflink{https://doi.org/} 

\Title{Resilience of LTE-A/5G-NR Links Against Transient Electromagnetic Interference}

\TitleCitation{Resilience of LTE-A/5G-NR Links Against Transient Electromagnetic Interference}

\Author{Sharzeel 
 Saleem\orcidA{}  $^{\dagger,\ddagger}$* and  Mir Lodro \orcidB{} $^{\dagger,\ddagger}$}

\AuthorNames{Sharzeel Saleem and Mir Lodro}

       \AuthorCitation{Saleem, S.; Lodro, M.}

\address{%
{School of Electrical, Electronic and Mechanical Engineering, University of Bristol, BS8 1TR} 
\\
}

\corres{Correspondence: sharzeel.saleem.2023@bristol.ac.uk}

\firstnote{Current address: {School of Electrical, Electronic and Mechanical Engineering, Queen's Building
University Walk, Bristol, BS8 1TR} 
}  
\secondnote{These authors contributed equally to this work.}

\abstract{This paper presents a comparative analysis of a long-term evolution advanced (LTE-A) and fifth-generation new radio (5G-NR), focusing on the effects of transient electromagnetic interference (EMI) caused by catenary--pantograph contact in a railway environment.A software-defined radio (SDR)-based prototype was developed to evaluate the performance of LTE-A and 5G-NR links under the influence of transient interference. The results show that both links experience considerable degradation due to interference at different centre frequencies. Performance degradation is proportional to the gain of interference. The measurement results show that both links experience considerable performance degradation in the presence of transient EMI.}


\begin{document}



\section{Introduction}\label{sec1}

Wireless systems such as 5G-NR and LTE-A operate in complex propagation environments and are susceptible to EMI, such as transients generated by high-power generators or emissions from loose pantograph--catenary systems, which degrades their performance \cite{de2023contribution,de2023designing}. This paper aims to establish the impact of transient EMI on the susceptibility of LTE-A/5G-NR communication links.

Transient EMI can cover a wide range of the spectrum, including Wi-Fi and cellular bands. Hence, EMI can affect both the primary user link from the eNB (evolved Node B)/gNB (next-generation Node B) to the user equipment (UE) onboard and on the platform. It can also affect the performance of secondary wireless links, such as WiFi access points (APs), to onboard users. It can also introduce additional problems in spectrum-sharing scenarios where primary and secondary user networks coexist and use energy detection to detect the presence of primary users. Transient EMI sources are difficult to locate. Furthermore, transient EMI can appear to be random and inconsistent, depending on the physical state of the fault location. The intermittent contact between the catenary and pantograph generates a wideband signal capable of jamming communication links across a broad range of frequencies. This paper aims to investigate the effects of unintentional interference on these communication links. 

The key performance indicators (KPIs) used to measure these effects are error vector magnitude (EVM), bit error rate (BER), and signal-to-interference-plus-noise ratio (SINR). We divided our work into five sections. Section \ref{sec1} is the introduction. In Section \ref{sec2}, we discuss the background of the problem and highlight the latest contributions. In Section \ref{sec3}, we introduce the setup of the experiment, our methodology and explain KPIs. Section~\ref{sec4} is about the measurement results and the effect of transient interference on LTE-A and 5G-NR. The conclusion of this study is given in Section \ref{sec5}.

\section{Literature Survey}\label{sec2}
Studies on transient EMI primarily concentrate on assessing the performance of communication links by evaluating KPIs. The authors in \cite{fan2023emi} have provided a comprehensive overview of the impact of intentional and unintentional sources of EMI on the global system for mobile communication (GSM), Long-Term Evolution-Railway (LTE-R), and {5G-Railway} 

 (5G-R). They have detailed the sources of EMI not only from pantograph arcing but also from various other sources of intentional and unintentional EMIs. They have explained the impact of interference from onboard power electronics and narrowband jammers. Their measurement results suggest a degrading impact on system performance, particularly concerning BERs as the interference power increases.

The work in \cite{fan2024deep} presents a real-time classification method for EMI and intentional EMI. The authors have provided a brief overview of transient and sweep interference. They have used machine learning to check the classification accuracy of different sources of interference at a centre frequency of 1.9 GHz and different channel parameters. The works in \cite{romero2017identication,deniau2017ieee} 
focus on the evaluation of intentional and unintentional interferences found in the transportation sector, particularly used in high-speed train scenarios. They have particularly focused on the interference effect on the IEEE {802.11n} 
 Wi-Fi hotspot onboard trains. The IEEE 802.11n standard operates in the 2.4 GHz and 5 GHz bands with different modulation and bandwidth requirements.

The authors in paper \cite{wang2019influence} have also evaluated the influence of pantograph arcing on the LTE-R. With the help of a simulation, they showed the impact of arcing on the block error rate and LTE throughput. The study concluded with evidence of system performance degradation.
Similarly, the authors in \cite{stienne2020assessment} have shown a transient impact of EMI on LTE communication. They have considered the EVM and peak-to-average power ratio (PAPR) as the main KPIs for performance evaluation. They have provided a detailed account of the LTE PHY layer and have shown the impact of EMI on the EVM and PAPR as the interference-to-signal power ratio increases.

Ref. \cite{romero2017evaluation} has evaluated the impact of pantograph--catenary transient emission on the performance of the IEEE 802.11n standard in the entire 2.4--2.5-GHz band. They have performed detailed bit error measurements with different time interval durations. They have found that the bit error becomes worse as the duration of the transient interference increases.

The work presented by \cite{correia2019analysis} includes a detailed study of the impact of pantograph--catenary emissions on railway communications. They have focused on the evaluation of antenna systems' deployment for GSM-R and LTE-R at frequencies of 380, 900, 2600, and 5900 MHz, respectively. 

The research presented by \cite{article_TEMI} examines electromagnetic disturbances caused by loose catenary--pantograph contact. The experiment demonstrated that when the pantograph--catenary spacing remains constant, increasing the voltage level leads to a higher current in the discharge circuit, which intensifies impulse radiation during contact loss events. Long-Term Evolution-Metro (LTE-M) is a dedicated communication system for train control, with stringent requirements regarding adjacent channel interference (ACI). The research in \cite{s22103876} proposes a solution using the Monte Carlo method to analyse the convergence of different scenarios where LTE-M and ACI coexist. 
A summary of the literature related to the concept is shown in Table \ref{tab:LR}.

\begin{table}[H]
\caption{Summary of literature review.\label{tab:LR}}
	\begin{adjustwidth}{-\extralength}{0cm}
		\begin{tabularx}{\fulllength}{CCCC}
\toprule
\textbf{Citation} & \textbf{Key Idea}  & \textbf{Impact} \\
\midrule
\cite{fan2023emi} & Jamming attacks in electrified railway environments. & Narrowband jammers have lethal and prolonged effects on communication links compared to UWB jammers.  \\
\bottomrule

\cite{fan2024deep}& Leveraging Bidirectional Long Short-Term Memory (BiLSTM) networks for sequential pattern learning and interference mitigation. & The BiLSTM-based model efficiently processes time-series signals, enabling real-time detection and classification with a 93.4\% accuracy, surpassing previous approaches. \\
\midrule
\cite{romero2017identication, deniau2017ieee} & Analysed OFDM signal vulnerability and the impact of interference parameters on communication degradation, considering reception-stage signal processing.  & The study showed that interference depends on the sweep period and receiver's time window. A longer observation window reduces the impact, affecting fewer subcarriers. Adjusting the window minimises interference. \\
\midrule
\cite{wang2019influence, stienne2020assessment} &  Transient EMI generated by catenary and pantograph contact impacts the LTE link. & The peak-to-average power ratio (PAPR) evaluation metric is proposed instead of the EVM because the EVM breaks down when the value exceeds 17.5\%. \\
\midrule
 \cite{romero2017evaluation} & Susceptibility study of an IEEE 802.11n network in a railway electromagnetic (EM) environment.  & There is a direct relationship between the interference power, its duration, and the highest average power measured by the Clear Channel Assessment (CCA), which affects the BER. \\
 \midrule
 \cite{correia2019analysis}  & Analysed antenna performance for four railway communication systems (TETRA at 380 MHz, GSM-R at 900 MHz, LTE-R at 2.6 GHz, and BBRS at 5.9 GHz) under the influence of the pantograph and catenary. &  The omnidirectional behaviour of the antenna was lost due to the presence of loose contact between the pantograph and catenary.\\
 \midrule
\cite{article_TEMI, s22103876} & Developed a test system for EMI caused by high-voltage discharge in a pantograph--catenary system. & Electromagnetic disturbances from loose catenary--pantograph contact increase with higher voltage, intensifying impulse radiation.  \\ \midrule
 \\ This article & Developed a system to test interference in LTE-A/5G-NR systems using USRPs. & 5G-NR is susceptible to interference from transient EMI. \\

\bottomrule
\end{tabularx}
\end{adjustwidth}
\end{table}

\section{Materials and Methods}\label{sec3}

This experiment evaluated the impact of transient EMI on a controlled communication link established using commercial off-the-shelf (COTS) software-defined radios from National Instruments. This study utilized the {National Instruments USRP-2943R (Europe),} 
which is functionally equivalent to the {Ettus Research X310 USRP,} 
 along with a {Ettus Research CBX-40 RF daughterboard (Europe) } 
 card providing full-duplex operation with a 40 MHz bandwidth.  

The experimental setup comprised three NI USRPs equipped with Ettus Research CBX-40 RF daughterboard (Europe)  RF cards. Two USRPs were used to establish the primary communication link by transmitting and receiving standardized LTE and 5G-NR waveforms, ensuring repeatable BER and EVM performance. The {LTE Toolbox} 
 and {5G Toolbox} were employed for waveform generation and processing, while {MATLAB version R2024b} 
 Wireless Testbench facilitated high-speed transmission and reception of complex waveforms without dropped samples.  

A third National Instruments USRP (Europe), with identical RF and digital specifications, was used to generate RF interference signals at various centre frequencies. A Rohde \& Schwarz spectrum analyser monitored the power spectral density (PSD) of the received signal in the presence of interference. Frequency scanning between 2.194 and 2.2045 GHz identified the most EMI-sensitive centre frequencies, with interference gains set at 12 dB, 15 dB, and 18 dB.  

The findings from the experiment highlighted a significant impact at an 18 dB gain, which effectively disrupted communication, underscoring the potential severity of transient EMI at higher interference gains. This effect of the 18 dB gain on signal degradation is discussed in detail in this paper, revealing critical insights into EMI resilience and vulnerability in different frequencies and gain settings.

\subsection{Types of Unintentional Interferences}
\begin{figure}[H]
    \includegraphics[width=12 cm, height=7.25 cm]{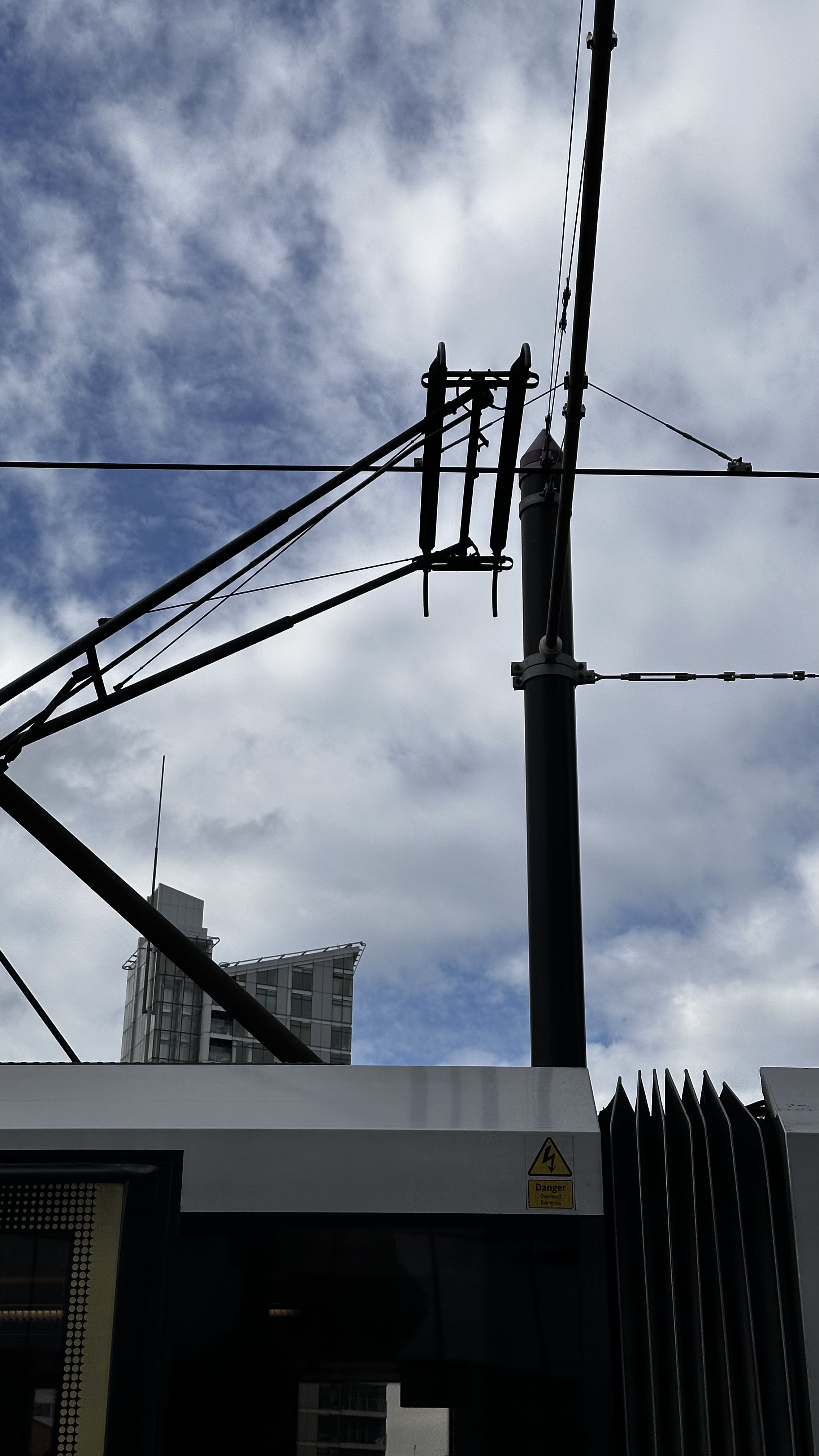}
    \caption{Catenary--pantograph contact.\label{fig:cp}}
\end{figure}

Unintentional interference such as transient interference can come from different sources. In the railway environment, unintentional transient interference is produced by high-power generators and the arcing effect of pantograph--catenary contact. Unintentional interference can arise from numerous sources, such as switching of power supplies or operation of high-power motors. A notable example is the EMI produced when pantographs on trains come into contact with catenary wires, as shown in Figure \ref{fig:cp}.

\begin{figure}[H]
    \includegraphics[width=12cm, height=7.25cm]{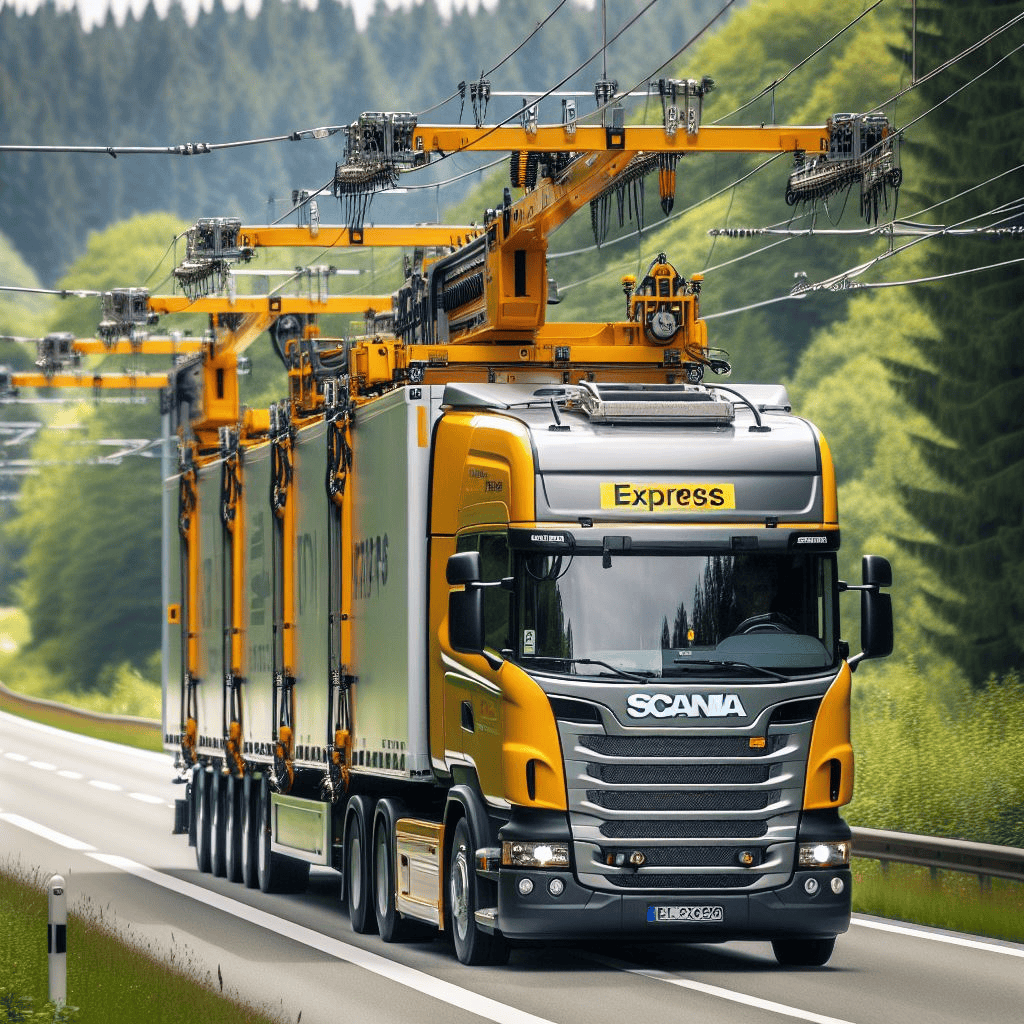}
    \caption{Road transport powered by electricity.}
    \label{fig:ELE}
\end{figure}

 Currently, innovations in power supply systems for road transport have aimed to reduce emissions, but they also introduce potential sources of transient EMI. As shown in Figure \ref{fig:ELE} (generated with the assistance of {{GenAI ChatGPT Turbo version 4.1)},} 
 the movement of vehicles on highways equipped with these systems can inadvertently interfere with communication links. The effect can be severe if tram overhead lines pass through the city centre or critical infrastructure, which can disrupt communication links such as Long Range (LoRA), Wi-Fi, and LTE-A \cite{romero2017identication}. Another source of unintentional interference is the SPN-43 radar, which operates in the S-band for the location of military targets, which can disrupt nearby communication systems. Furthermore, dynamic spectrum sharing (DSS) in the citizens broadband radio service (CBRS) spectrum improves spectrum efficiency, but it can also introduce interference between shared bands, challenging communication reliability \cite{fan2023emi}.

\subsection{Testbed Configuration and Parameter Specifications}
\begin{figure}[H]
    \includegraphics[width=12cm]{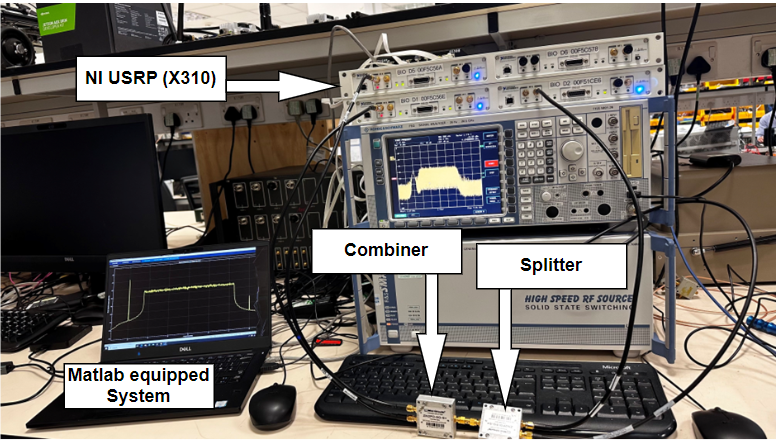}
    \caption{Photo of the experimental setup.}
    \label{fig:ES}
\end{figure}

The experimental setup facilitates a thorough analysis of how various interference parameters affect the stability and performance of the communication link. In the context of an LTE-A/5G-NR communication link, a series of experiments were carried out. The PHY layer parameters of the LTE-A and 5G-NR are shown in Tables \ref{tab:lte-param} and \ref{tab:5g-param}{,} 
 respectively. For the transmission and reception of the baseband waveforms, we used separate USRPs to rule out the effect of cross-channel leakage. Figure \ref{fig:ES} shows the photo of the experimental setup where three { National Instruments X310 USRPs (Europe)} 
 are connected to a { laptop with an Intel Core i7 processor and at least 32 GB of RAM.} 
 via an Ethernet switch. We generate transient EMI at various RF frequencies within the bandwidth of interest. The initial phase included frequency scanning to pinpoint the frequencies most susceptible to performance degradation under jamming conditions. The findings revealed that the centre frequencies of 2.1955 GHz, 2.2 GHz, and 2.2045 GHz were especially vulnerable to interference.

\begin{table}[H] 
\caption{LTE-A PHY communication link parameters.\label{tab:lte-param}}
\begin{tabularx}{\textwidth}{CC}
\toprule
\textbf{Parameter} & \textbf{Value}  \\
\midrule
Base RMC Configuration & R.7 (10 MHz)  \\
Jammer Sampling Rate & 15.36 MHz  \\
Power Scale Factor & 0.8  \\
Jammer Radio Gain & 2 dB  \\
eNodeB (TX) Radio Gain & 25 dB  \\
UE (RX) Radio Gain & 20 dB  \\
Center Frequency & 2.2 GHz  \\
EVM Peak & 17.771\%  \\
EVM RMS & 5.049\%  \\
BER & 0 \\
Frequency Offset & 1372.998 Hz  \\
\bottomrule
\end{tabularx}
\end{table}
The bandwidth used in this experiment is 10 MHz. Additionally, the EVM is worse as the interference increases, particularly for higher-order modulation such as 64-QAM. In the absence of interference, it was observed that the EVM RMS is approximately 5\%, which is as low as possible with a zero BER. The generated waveform uses 64-QAM modulation with a forward error correction (FEC) code rate of 3/4, a subcarrier spacing of 30~kHz, and a bandwidth of 10 MHz, and it supports up to 24 resource blocks. 

\begin{table}[H] 
\caption{5G-NR communication link parameters.\label{tab:5g-param}}
\begin{tabularx}{\textwidth}{CC}
\toprule
\textbf{Parameter} & \textbf{Value} \\
\midrule
Frequency Range & 410 MHz--7.125 GHz \\
Modulation Coding Scheme & 64-QAM, R = 3/4 \\
Channel Bandwidth & 10 MHz \\
Sampling Rate & 15.36 MHz \\
TX Radio Gain & 20 dB \\
RX Radio Gain & 20 dB \\
Jammer Radio Gain & 2 dB \\
Center Frequency & 2.2 GHz \\
Subcarrier Spacing & 30 kHz \\
\bottomrule
\end{tabularx}
\end{table}

\subsection{ 
	Transient Electromagnetic Interference (EMI) and Unintentional Interference}
The interaction between the catenary and pantograph generates transient broadband electromagnetic waves. Numerous studies have been conducted on the characterization of these fast-radiated transient disturbances resulting from the sliding contact of the pantograph on the catenary. These studies utilize various approaches, including a spectral approach, as discussed in article \cite{tr-1}, a temporal approach, as explored in article \cite{article-tr-2}, and a combination of both approaches, as examined in articles \cite{6903922-t3,inproceedings-tr4}. The setup utilizes a double-sided transient waveform to investigate the effects on the LTE communication link. In the RailCom project \cite{article-tr-2,article_2_ep1, 5399370_eq_1_ar_3}, measurement campaigns were conducted on high-speed trains to assess the emissions received by the GSM-R antenna, which were caused by catenary--pantograph contact losses. To examine the influence of fast transient signals, the signal was modulated at the studied RF frequency, effectively confining the transient interference effect to the targeted bandwidth. The statistical analysis of fast transient interferences \cite{tr-1} yielded rise times and hold times within specific ranges.

\begin{figure}[H]
    \includegraphics[width=12cm]{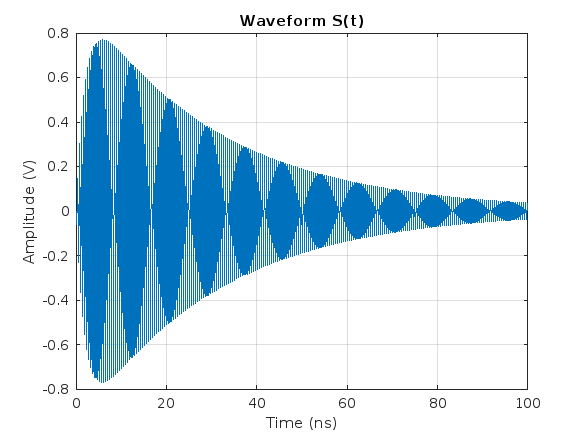}
    \caption{{Double$-$sided} 
 exponential wave at 2.2 GHz.}
    \label{fig:ds}
\end{figure}

 Mathematically, a wave can be written as a double-sided transient waveform and is given by
\begin{equation}
s_{\text{trans}}(t) = A_0 + e^{-\frac{t}{\tau_{\text{rise}}}} + e^{-\frac{t}{\tau_{\text{hold}}}} + \sin(2 \pi f_c t) + u(t)
\end{equation}
where $A_0$ is the amplitude; $f_c$ is the carrier frequency; $u(t)$ is the unit step function; $\tau_{\text{rise}}$ is the rise time constant, which can be in the range of 0.1 and 3 ns but was taken as 2 ns for this experiment; and $\tau_{\text{hold}}$ is the hold time constant, which can be in the range of 1 ns and 50 ns but was taken as 30 ns for this experiment. Figure \ref{fig:ds} illustrates the double-sided exponential waveform generated using the aforementioned equation, which serves as a means of interference.

\subsection{Key Performance Indicators}
The primary KPIs are the EVM and SINR.

\subsubsection{Error Vector Magnitude (EVM)} 
The EVM is defined as the measure of how far the actual transmitted signal deviates from the ideal or reference signal in a constellation diagram. In simpler terms, it measures the difference between the ideal signal points and the points where the actual received signal ends up due to various imperfections in the transmission process \cite{stienne2020assessment}. The equation for calculating the EVM is given by

\begin{equation}
\text{EVM} (\%) = \frac{\sqrt{\frac{1}{N} \sum\limits_{i=1}^{N} \left| S_\text{ref}[i] - S_\text{received}[i] \right|^2}}
{\sqrt{\frac{1}{N} \sum\limits_{i=1}^{N} \left| S_\text{ref}[i] \right|^2}} \times 100
\end{equation}

\noindent where $S_\text{ref}[i]$ is the ideal symbol or the reference symbol at the $i$-th symbol and $S_\text{received}[i]$ is the $i$-th received symbol. $N$ is the total number of complex symbols.

\subsubsection{Signal-to-Interference-Plus-Noise Ratio (SINR)}

The SINR evaluates the quality of the radio link between a base station and the UE. It is the ratio of the power of the intended received signal, $P_\text{signal}$, to the power of the interference signal, $P_\text{interference}$, plus the noise power, $P_\text{noise}$:
\begin{equation}
\text{SINR} = \frac{P_\text{signal}}{P_\text{interference} + P_\text{noise}}
\end{equation}

For OFDM-specific waveforms, the SINR at the $k$ subcarrier  is
\begin{equation}
\text{SINR}_k = \frac{|H_k|^2 P_\text{Tx}}{\sum_{j \neq k} |I_j|^2 P_\text{Tx} + N_0 B_k}
\end{equation}
where $H_k$ is the channel frequency response for the $k$ subcarrier, $P_\text{Tx}$ is the power of the signal transmitted per subcarrier, $|I_j|^2$ is the power of interference, and $N_0$ and $B_k$ are the spectral density of the noise power and the bandwidth of the $k$ subcarrier, respectively. 

We consider the signal power received from the demodulated OFDM symbols. This metric is crucial for network operators, as it helps assess and optimize network resilience.


\section{Results and Discussions}\label{sec4}
The experimental results are organized into two sections: The first section details the impact of a double-sided exponential jammer (Transient EMI) on an LTE-A link. In contrast, the second section examines its effect on a 5G-NR link. Finally, a comparison is performed to determine which communication link is more susceptible to this type of interference.


\subsection{Transient EMI in LTE-A}

A double-sided exponential jammer with adjustable gain and varying centre frequencies was tested. The impact was substantial at a transmit gain of 18 dB, with KPIs being summarized in Table  \ref{tab:Risk51} below. The findings from the experiment demonstrate that LTE-A link performance is highly dependent on the interferer frequency, with 2.2 GHz exhibiting the most severe impact, as reflected by the highest EVM peak (1985.256\%), EVM RMS (85.549\%), and BER (0.32059). This is likely due to increased spectral overlap with the desired signal, leading to significant degradation. In contrast, at 2.1955 GHz, the EVM peak (829.12\%) and BER (0.24165) are lower, indicating better resilience to interference. The observed frequency offset (2132.578 Hz at 2.2 GHz vs. 1680.86 Hz at 2.1955 GHz) further supports the conclusion that certain frequencies are more vulnerable to transient EMI.

\begin{table}[H] 
\caption{ KPI measurements of the  LTE-A link at different RF frequencies in the presence of transient interference. The gain of the interfering USRP is set to 18 dB.}\label{tab:Risk51}
\begin{tabularx}{\textwidth}{CCCCCC}
\toprule
\textbf{Interferer Frequency} & \textbf{EVM Peak (\%)} & \textbf{EVM RMS (\%)} & \textbf{BER} & \textbf{Frequency Offset (Hz)} & \textbf{SINR (dB)} \\
\midrule
2.1955 GHz & 829.12 & 79.212 & 0.24165 & 1680.86  & 1.89 \\
2.2 GHz & 1985.256 & 85.549 & 0.32059 & 2132.578  & 4.5046 \\
2.2045 GHz & 1387.908 & 83.565 & 0.26159 & 2082.8  & $-$0.2355 \\
\bottomrule
\end{tabularx}
\end{table}

The experimental results indicate a significant degradation in the SINR, which decreased from 15.1226 dB (no interference)  to  $-$0.2355 dB. This drop represents the maximum performance degradation observed in the LTE communication link, despite employing a high transmit gain intended to mitigate such losses.

\begin{figure}[H]
    \includegraphics[width=12cm, height=7cm]{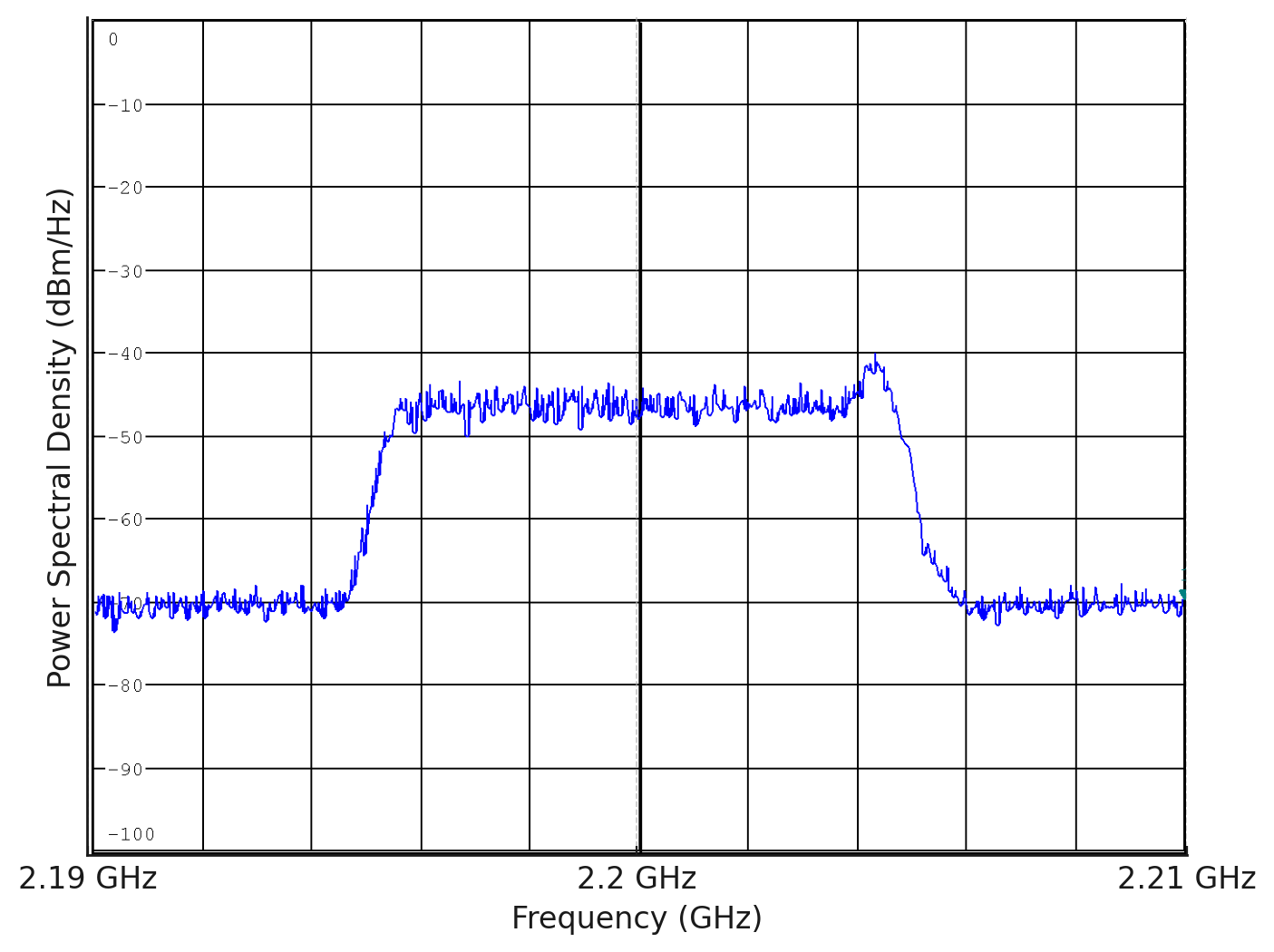}
    \caption{{Power} 
 spectral density due to EMI at 2.2045 GHz.}
    \label{fig:wb-2.2}
\end{figure}

\begin{figure}[H]
    \includegraphics[width=12cm,height=7cm]{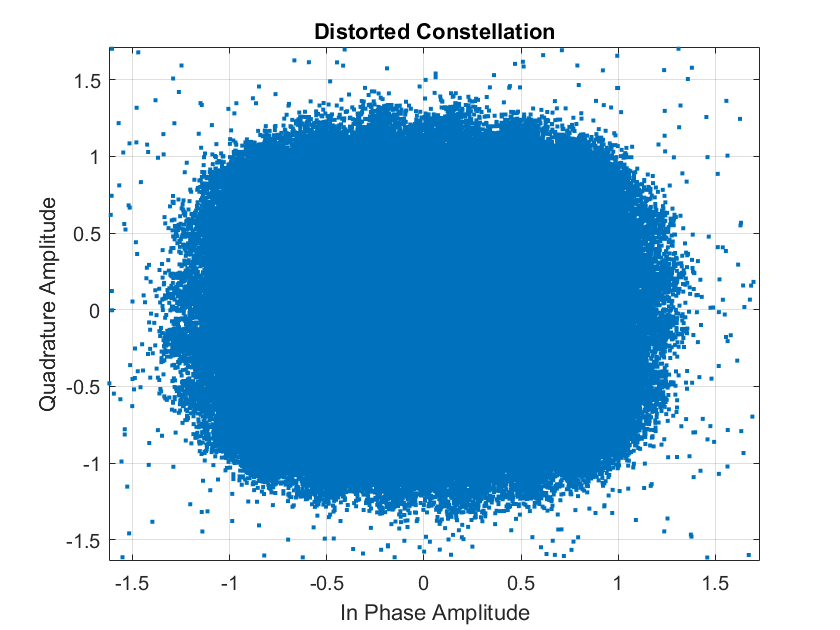}
    \caption{{Constellation} 
 due to jamming at 2.2045 GHz.}
    \label{fig:cons}
\end{figure}

\begin{figure}[H]
    \includegraphics[width=12cm,height=7cm]{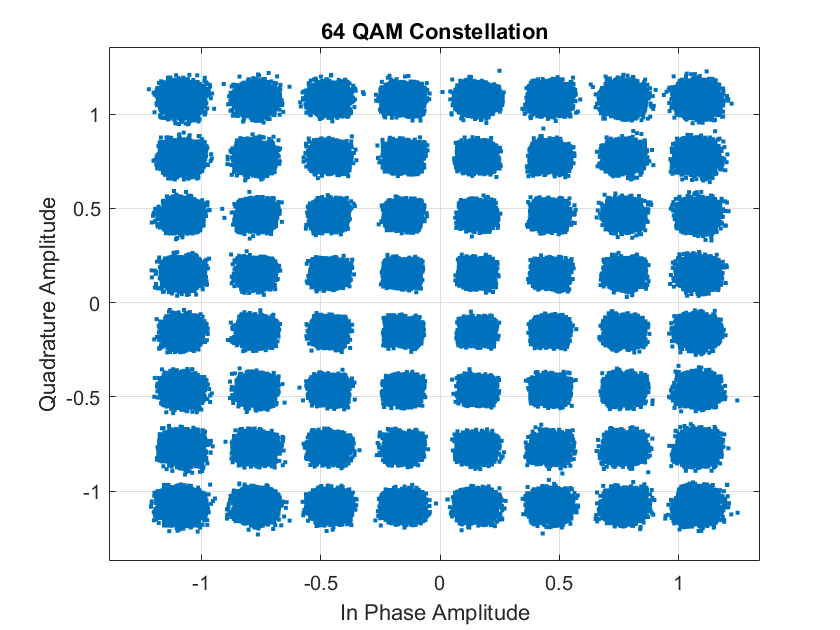}
    \caption{{Received} 
 64$-$QAM constellation.}
    \label{fig:e}
\end{figure}

Figure \ref{fig:wb-2.2} illustrates a peak centred at 2.2045 GHz, resulting from the injection of transient EMI. This EMI not only contributes to the SINR reduction but also leads to substantial distortion in the signal constellation. As shown in {Figure} 
 \ref{fig:cons}, the constellation exhibits the effect of noise that can vary between symbols. This noise around the ideal symbol configuration suggests that the received symbols are displaced from their intended positions, reflecting the detrimental effects of frequency distortion on the integrity of the signal and the overall performance of the LTE communication system. This can be verified by comparing Figure \ref{fig:e}, which a perfectly received 64 QAM constellation, with Figure \ref{fig:cons}, showing distortion. Additionally, the received constellation appears to be a single noise cloud around the origin if the magnitude of the interference increases to a level where the decodability of the data symbols is not possible. These findings underscore the critical impact of transient EMI on the reliability of LTE communications.

\subsection{Transient EMI in 5G-NR}

A double-sided exponential signal was introduced into the primary communication link at three distinct transmit gains, mirroring the configuration used in the LTE link. All parameters were kept constant, except for the operating frequency. As indicated in the table, the signal at 2.2 GHz shows the greatest performance degradation, although its impact is still less severe compared to that of the continuous wave (CW) tone jammer. Table~\ref{tab:Ris} demonstrates that the 5G-NR link's performance varies significantly across different interferer frequencies, with 2.2 GHz showing the most severe impact in terms of the BER (0.3884) and SINR ($-$12.67 dB), which is the lowest. This frequency likely causes the greatest disruption because of the strong spectral overlap with the desired signal, leading to higher bit errors and a reduction in the signal quality. The exceptionally high EVM values at 2.1955 GHz (EVM peak: 389223.16\%) suggest extreme signal distortion, even though its BER is lower than at 2.2 GHz. This indicates that while interference at 2.1955 GHz causes significant distortion, it may not cause as much bit error rate degradation, possibly due to the frequency's reduced interaction with the signal's key components.

\begin{table}[H] 
\caption{KPI measurements of the 5G-NR link at different RF frequencies in the presence of transient interference. The gain of the interfering USRP is set to 18 dB.\label{tab:Ris}}
\centering
\begin{tabularx}{\textwidth}{CCCCC}
\toprule
\textbf{Interferer Frequency} & \textbf{EVM Peak (\%)} & \textbf{EVM RMS (\%)} & \textbf{BER} & \textbf{SINR (dB)} \\
\midrule
2.1955 GHz & 389223.16 & 3201.15 & 0.3214 & 3.52 \\
2.2 GHz & 107336.92 & 1774.27 & 0.3884  & $-$12.67 \\
2.2045 GHz & 285892.70 & 2446.83 & 0.3503 & $-$3.64 \\
\bottomrule
\end{tabularx}
\end{table}

\begin{figure}[H]
    \includegraphics[width=12cm, height=7cm]{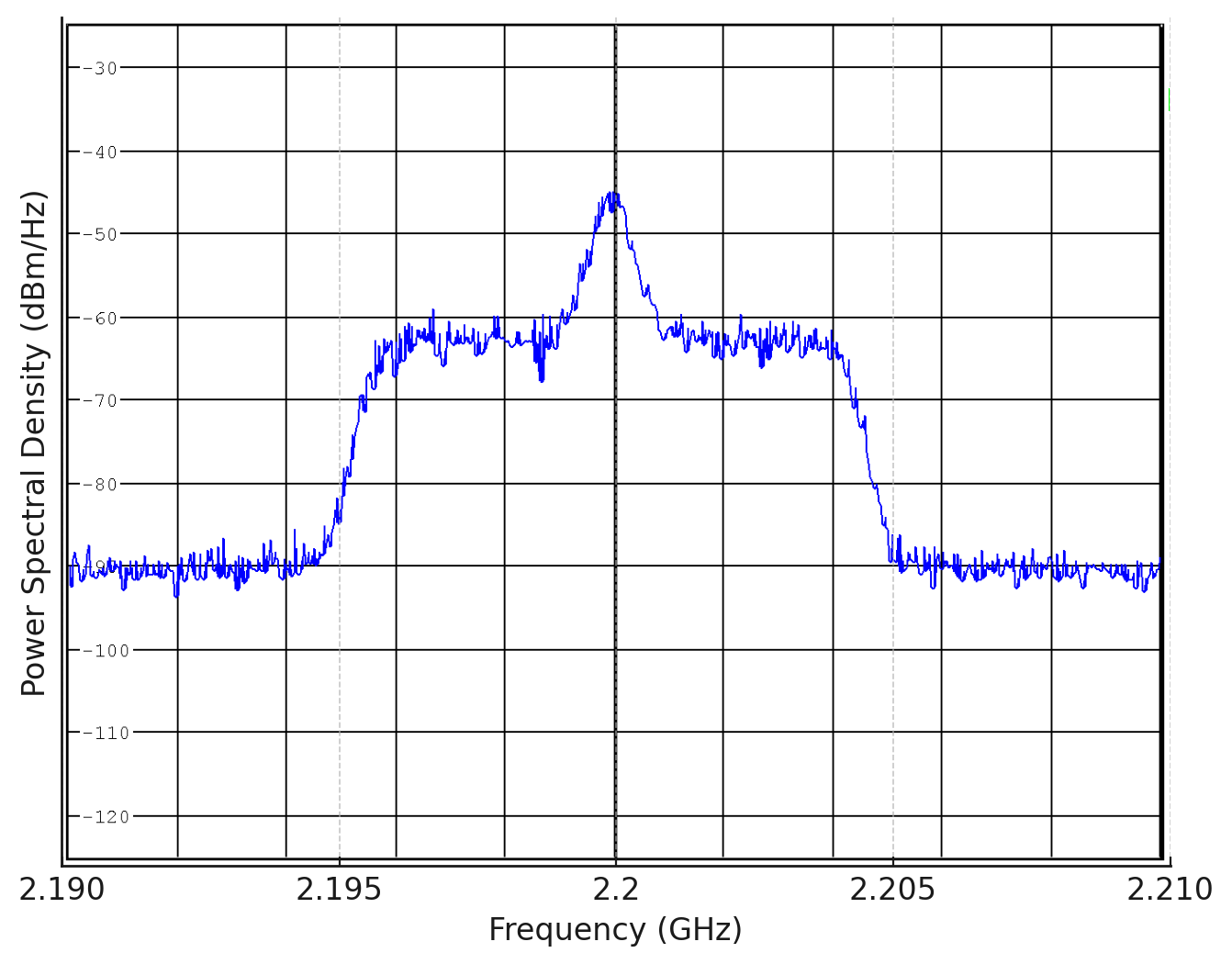}
    \caption{{Power} 
 spectral density due to EMI at 2.2 GHz.}
    \label{fig:spectrum_2.2}
\end{figure}

\begin{figure}[H]
    \includegraphics[width=12cm]{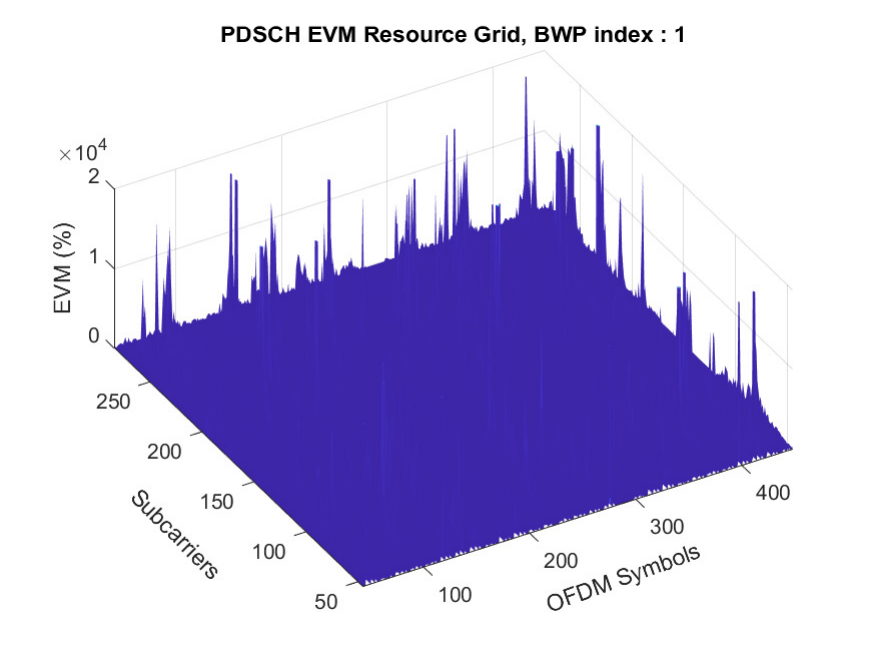}
    \caption{EVM resource grid at 2.2 GHz under interference.}
    \label{fig:EVM_2.2GHz}
\end{figure}

\begin{figure}[H]
    \includegraphics[width=12cm]{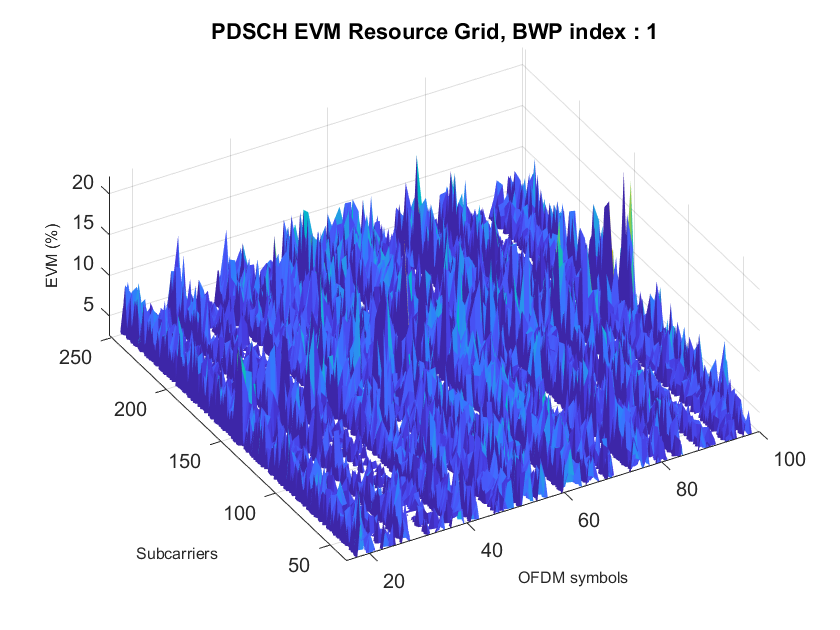}
    \caption{EVM resource grid without interference.}
    \label{fig:dk}
\end{figure}

A distinct peak can be observed in Figure \ref{fig:spectrum_2.2}, resulting from the jammer effect. This peak indicates a significant increase in energy, which consequently distorts the EVM resource grid shown in {Figure} 
 \ref{fig:EVM_2.2GHz}. This effect becomes evident when the jammer's transmit gain is set to 18 dB. The presence of the jammer induces substantial interference in the communication link, leading to a higher EVM peak. A comparison can be made by referring to Figure \ref{fig:dk}, where a considerable number of OFDM symbols and subcarriers exhibit a low EVM\%. 
 
The high EVM values indicate a degradation in signal quality, reflecting the extent of distortion caused by the interference signal. As the interference gain increases, the interference becomes more pronounced, negatively affecting the integrity of the transmitted signal. This scenario highlights the critical impact of jamming on communication systems, particularly in environments where the SNR is already compromised. The distortion in the EVM resource grid further illustrates the challenges in maintaining reliable communication under such adverse conditions, underscoring the necessity for robust interference mitigation strategies in practical implementations.

The final EVM and BER evaluation of both communication links is summarized in Table \ref{tab:overall}. It can be observed that the BER and EVM in both communication links are high when there is interference at the centre frequency of 2.2045 GHz at an interference gain value of 18 dB.

\begin{table}[H]
\caption{Final measurement results of LTE-A and 5G-NR communication links with and without interference.\label{tab:overall}}
	\begin{adjustwidth}{-\extralength}{0cm}
		\begin{tabularx}{\fulllength}{CCCC}
			\toprule
			\textbf{Interferer Type} & \textbf{EVM Peak (\%)} & \textbf{BER (\%)} & \textbf{SINR (dB)} \\
			\midrule
\multirow{1}{*}{LTE-A Link (No Interference)} & 17.77 & 0.00 & 15.12 \\  \midrule
\multirow{1}{*}{5G-NR Link (No Interference)} & 23.76 & 0.00 & 15.97 \\  
\midrule
\multirow{1}{*}{LTE-A with Transient EM Wave \textsuperscript{1}  } & 1387.91 & 0.26 & $-$0.23 \\\midrule
\multirow{1}{*}{5G-NR with Transient EM Wave \textsuperscript{2} } & 107336.97 & 0.39 & $-$12.67 \\  
\bottomrule
		\end{tabularx}
	\end{adjustwidth}
\noindent{\footnotesize{\textsuperscript{1} Jammer @ 2.2045 GHz, 18 dB Gain. \textsuperscript{2}  Jammer @ 2.2 GHz, 18 dB Gain.}}

\end{table}

\section{Conclusions and Future Scope}\label{sec5}
This study presents an experimental evaluation of communication link susceptibility using software-defined radio platforms. The results indicate that 5G-NR links experience greater degradation in capacity and the bit error rate (BER) under unintentional interference. Previous studies have reported disruptions near hospitals and universities, particularly in areas with tram routes. The findings further support the claim that LTE systems exhibit greater resilience to high-frequency interference.

The primary objective of this work was to analyse performance degradation caused by this type of interference.  
Our future research will focus on assessing its impact when combined with a narrowband smart jammer, with the aim of developing mitigation strategies to address this issue.

\authorcontributions{Sections \ref{sec1}--\ref{sec5}, S.S.; supervision and project administration, M.L.
All authors have read and agreed to the published version of the manuscript.}

\funding{This research received no external funding.}

\institutionalreview{Not applicable.}

\informedconsent{Not applicable.}

\dataavailability{Data are unavailable to due privacy and ethical restrictions.} 

\acknowledgments{The authors express their gratitude to the University of Bristol for providing the support and resources that made this research possible. During the preparation of this manuscript/study, the authors used [{GPT-4-turbo version 4.1}] for the purposes of generating Figure \ref{fig:ELE}. The authors reviewed and edited the output and take full responsibility for the content of this publication.}

\conflictsofinterest{The authors declare no conflicts of interest.} 

%


\begin{adjustwidth}{-\extralength}{0cm}
\reftitle{References}

\PublishersNote{}


\end{adjustwidth}

%



\end{document}